# $\mathcal{G}$-SELC: OPTIMIZATION BY SEQUENTIAL ELIMINATION OF LEVEL COMBINATIONS USING GENETIC ALGORITHMS AND GAUSSIAN PROCESSES

By Abhyuday Mandal,[1] Pritam Ranjan[2] and C. F. Jeff Wu[3]

*University of Georgia, Acadia University and Georgia Institute of Technology*

Identifying promising compounds from a vast collection of feasible compounds is an important and yet challenging problem in the pharmaceutical industry. An efficient solution to this problem will help reduce the expenditure at the early stages of drug discovery. In an attempt to solve this problem, Mandal, Wu and Johnson [*Technometrics* **48** (2006) 273–283] proposed the SELC algorithm. Although powerful, it fails to extract substantial information from the data to guide the search efficiently, as this methodology is not based on any statistical modeling. The proposed approach uses Gaussian Process (GP) modeling to improve upon SELC, and hence named $\mathcal{G}$-SELC. The performance of the proposed methodology is illustrated using four and five dimensional test functions. Finally, we implement the new algorithm on a real pharmaceutical data set for finding a group of chemical compounds with optimal properties.

**1. Introduction.** Finding global optima of complex physical processes with large input spaces (or search spaces) is one of the primary goals in many scientific investigations. For example, scientists in pharmaceutical industries can often produce a large number of compounds. However, only a few of them would serve as good candidates for a potential drug. While compounds may be designed to be theoretically effective, their interactions with different parts of the body (e.g., liver, kidney, intestine, etc.) often render them ineffective or toxic [Welling (1997)]. Recent technological advancement has now enabled chemists to explore a large number of new potential compounds in a short period of time. The technology, known as

Received October 2007; revised August 2008.
[1]Supported by grants from the University of Georgia Research Foundation.
[2]Supported by Acadia University Research Grant.
[3]Supported by NSF Grants DMI-06-20259 and DMS-07-05261.
*Key words and phrases.* Process optimization, batch-sequential design, expected improvement function, Kriging.

combinatorial chemistry, is now being widely applied in the pharmaceutical industry, and is gaining interest in several areas of chemical industry as well [Lam, Welch and Young (2002), Leach and Gillet (2003), Gasteiger and Engel (2003)]. Combinatorial chemistry uses robotics to combine sets of monomers to create thousands of new compounds at a time. In a typical compound, a core molecule is identified to which monomers are attached at multiple locations. Each attachment location may have tens or hundreds of potential monomers. Clearly, the compound library (set of all structurally feasible compounds) can become dauntingly large for a core molecule with just a few attachment points. This technology has been used in pharmaceutical industries to enhance the diversity of compound libraries, and to optimize endpoints such as target efficacy or one (or more) of the ADMET (absorption, distribution, metabolism, excretion, toxicology) properties of compounds [Rouhi (2003)]. Constrained by resources, most compound libraries cannot be fully synthesized and, thus, it is preferred to find smaller subsets of the libraries that consist of compounds with good desirable features.

In practice, pharmaceutical industries frequently use *ad hoc* methods based on the scientists' prior knowledge and intuition to find subsets of compound libraries that are small in size and contain compounds with desirable properties. In Section 2 such an instance, which motivated the current work, is discussed in more details. In this application the scientists used a four-stage procedure to create a subset of compounds with high process values (the exact chemical property of the response was not reported due to propriety reasons). Recently, Mandal et al. (2006) developed a Genetic Algorithm (GA)-based search procedure called Sequential Elimination of Level Combinations (SELC) for this purpose, which was motivated by the SEL algorithm of Wu, Mao and Ma (1990). SELC uses forbidden array and weighted mutation to enhance the performance of the search procedure compared to a standard GA [Holland (1975, 1992)]. Mandal et al. (2006) used several examples and real applications to illustrate that SELC outperforms classical GAs. Nonetheless, the lack of substantial statistical modeling in SELC leaves room for improvement.

In this paper we propose using the Gaussian Process (GP) modeling technique [see Sacks et al. (1989)] for developing a sequential search method ($\mathcal{G}$-SELC). As illustrated in Sections 5 and 6, the new approach outperforms SELC for the motivating pharmaceutical application and the simulated examples considered in the paper. $\mathcal{G}$-SELC is inspired by the works of Jones, Schonlau and Welch (1998), and Mandal et al. (2006). In their pioneering work, Jones et al. (1998) proposed a complete sequential (one trial at a time) sampling approach for process optimization by maximizing a merit based criterion called an expected improvement (EI) function. It turns out that restrictions on the compounds manufacturing process often requires chemical compounds to be manufactured in batches. That is, complete sequential approaches are undesirable for creating new promising compounds. Thus, we propose to select batches of trials using an adaptive mixture of SELC and EI. Four and five dimensional test functions are considered to illustrate the performance of the proposed approach versus the existing approaches.

The paper is organized as follows. The motivating combinatorial chemistry problem of a pharmaceutical company is presented in Section 2. Then, a brief review of the two existing methodologies, SELC and EI, are presented in Section 3. In Section 4 we develop the $\mathcal{G}$-SELC technique. Section 5 presents a comparison on the performance of the new methodology with that of the SELC and EI-based approach. In Section 6 $\mathcal{G}$-SELC is used for identifying potentially good compounds in the pharmaceutical industry example. Some concluding remarks are given in Section 7.

**2. Pharmaceutical chemistry example.** Identifying promising compounds from a vast collection of feasible compounds is a challenging problem in drug discovery [see "Finding Needles" in *Drug Discovery News*, Willis (2007)]. This research is motivated by a combinatorial chemistry problem, where the goal is to obtain sets of reagents (or monomers) that maximize the target efficacy of a compound, which is measured by its pre-specified physiochemical property for a specific biological screen.

Consider the data discussed by Mandal et al. (2007), where a compound was created by attaching reagents to the three locations denoted by $A$, $B$ and $C$ (core) of a molecule (see Figure 1). The data were obtained from a pharmaceutical company. In this application the compound library (the set of all feasible compounds) consisted of 5 feasible substructures (monomers) at position $A$, 34 at position $B$ and 241 substructures at position $C$. That is, the compound library had a total of 40,970 chemical compounds. Manufacturing all of these compounds was expensive, and thus, it was desirable to select a much smaller subset of the compounds with desirable properties. Once such a list of promising compounds was identified, the pharmacists synthesized them in the laboratory, and then the compounds with high responses were chosen for further analysis.

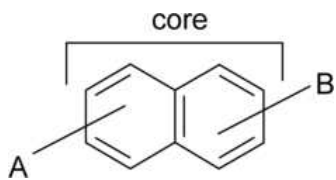

Fig. 1. *The core molecule of a compound with three reagents locations.*

4          A. MANDAL, P. RANJAN AND C. F. J. WUTABLE 1
*Characteristics of the subsets of the compound library chosen in different iterations of the optimization procedure: each column represents the substructures explored in each iteration and each row gives the number of new structures that were not examined in previous rounds*

| Iteration | $A$ | $B$ | $C$ | $A \times B$ | $A \times C$ | $B \times C$ | $A \times B \times C$ |
|---|---|---|---|---|---|---|---|
| 1 | 5 | 26 | 164 | 75 | 286 | 1168 | 2114 |
| 2 | 0 | 7 | 9 | 17 | 18 | 102 | 208 |
| 3 | 0 | 1 | 41 | 1 | 41 | 120 | 128 |
| 4 | 0 | 0 | 27 | 0 | 27 | 32 | 33 |
| Overall | 5 | 34 | 241 | 93 | 372 | 1422 | 2483 |

To explore the compound space, the scientists of the pharmaceutical company performed a 4-stage experiment. An initial subset of 2114 compounds was chosen on the basis of their scientific knowledge and intuition. These compounds were then created and screened (i.e., the response $y$ was obtained). Using the results from this initial screen and scientific knowledge about the target, three subsequent subsets of the compound library were generated. In total, 2483 compounds (6% of the possible compounds) were generated. Table 1 summarizes the substructure space explored at each iteration, and provides insight into the historical process optimization technique.

In the first iteration 2114 compounds were created. Looking at the structures of those compounds, one would observe that in the first iteration all the 5 substructures were explored for position $A$, 26 out of 34 substructures for position $B$, and 164 out of 241 substructures were explored for position $C$. Out of all possible 170 (=5×34) combinations of $A \times B$, only 75 distinct combinations occurred in the compounds of iteration 1. Other columns of Table 1 can be interpreted similarly. Note that the column corresponding to $A \times B \times C$ represents the actual number of compounds created in a particular iteration. In the second iteration no new substructures were explored for position $A$ (all were explored in the first iteration), but 7 new substructures for position $B$ and 9 new substructures for position $C$ were explored.

TABLE 2
*Desired compound characteristics for combinatorial chemistry example*

| | |
|---|---|
| Reactive matched | ¿1 |
| Risky matched | ¿3 |
| Molecular weight | ¿500 |
| Rotatable bonds | ¿10 |
| Rule of 5 | ¿2 |
| Aromatic ring count | ¿5 |
| Polar surface area | ¿140 |
| $\log P$ | ¿5 |

Compounds with process value greater than 10 (i.e., $y > 10$) were considered active (or good). Although not modeled in this paper, the active compounds that satisfy the constraints outlined in Table 2 were of more importance to the scientists. These constraints include the chemical properties: chemical reactivity, occurrence of toxicologically risky chemical features, molecular weight, number of rotatable bonds, violations of the Rule of 5 [Lipinski et al. (1997)], aromatic rings, calculated polar surface area and LogP (hydrophobicity). Of the 2483 compounds that were created, only 69% of the compounds satisfied the constraints in Table 2. The rest of the compounds had one or more undesirable characteristics. As a result, the number of active compounds found in the subset of size 2483 that met all desired characteristics was low.

From an experimenter's viewpoint, each compound can be thought of as a design point (run) and the different reagents as levels of the factors (i.e., positions). The physiochemical property of interest (e.g., percent inhibition values or any of the ADMET properties) of a compound can be viewed as the univariate response of the process, which is to be maximized. That is, the $i$th input (with $d$ independent factors) and the output of the underlying process evaluated at $n$ design points can be denoted by a $d$-dimensional vector, $x_i = (x_{i1}, \ldots, x_{id})$, and a scalar, $y_i = y(x_i)$, respectively. Although we consider only scalar valued response variables in this paper, multi-dimensional response (e.g., optimization of more than one ADMET properties) can also be accommodated by modeling the desirability scores [Mandal et al., (2007)] of the compounds.

In such experiments, one or more factors are often qualitative by nature, although in this paper we treat them as quantitative ones. Since the monomers (or reagents) can be ranked based on their intrinsic properties (viz. molecular weight, hydrophobicity or even abundance), such a simplification is justified in this setup. Hence, the proposed algorithm, $\mathcal{G}$-SELC, treats all the factors as quantitative (see Section 7 for more discussions). Next, we review the two key components of the $\mathcal{G}$-SELC algorithm.

**3. Review of existing methodologies.** The SELC algorithm developed in Mandal et al. (2006) is a GA-based optimizer that attempts to find subsets of the compound library which consist of good compounds. The EI approach proposed in Jones et al. (1998) is a complete sequential design strategy that is based on maximizing a figure of merit called the expected improvement (EI).

3.1. *SELC algorithm.* The SELC algorithm was proposed as an extension of SEL [Wu et al. (1990)] for finding optima of sufficiently undulating objective functions in a high-dimensional search space. This methodology was based on classical GAs, and did not use any model based information to search for the compounds with high process value (e.g., percent inhibition values, any of the ADMET properties or desirability scores). The features of the SELC algorithm that allow the algorithm to find optima quickly are the forbidden array and weighted mutation scheme.

TABLE 3
*The factor settings of the nine compounds with their process value*

| A | B | C | $y$ |
|---|---|---|-----|
| 1 | 1 | 1 | 10.1 |
| 1 | 2 | 2 | 53.6 |
| 1 | 3 | 3 | 43.8 |
| 2 | 1 | 2 | 13.4 |
| 2 | 2 | 3 | 46.9 |
| 2 | 3 | 1 | 55.1 |
| 3 | 1 | 3 | 5.7 |
| 3 | 2 | 1 | 43.6 |
| 3 | 3 | 2 | 47.0 |

*Forbidden array.* A collection of design points that have demonstrated poor fitness values, or a priori known to produce undesirable responses, is said to form a *forbidden array*. Such an array can be characterized by its *strength* and *order*. A forbidden array of strength $s$ consists of the worst $s$ runs of the experiment at each iteration of the algorithm. A forbidden array is said to be of order $k$ if any combination of $k$ or more levels from any design point in the forbidden array is prevented from being constructed in subsequent iterations of the algorithm. For instance, suppose three monomers (denoted by 1, 2 and 3) can be added to each of the three locations $A$, $B$ and $C$ of a core molecule as shown in Figure 1. Further suppose that only 9 compounds are created and analyzed (see Table 3). The fourth column, denoted by $y$, represents the process values of the 9 synthesized compounds. The forbidden array of strength 2 for this design consists of the 2 compounds that have the lowest $y$ values, namely, (3 1 3) and (1 1 1). That is, the following compounds will be prohibited: $\{(1\ 1\ *), (1\ *\ 1), (*\ 1\ 1), (3\ 1\ *), (3\ *\ 3), (*\ 1\ 3)\}$, where $*$ can take any admissible value.

*Weighted mutation.* The second main feature that makes the SELC algorithm unique is its *weighted mutation* scheme. After constructing the forbidden array, SELC searches for promising compounds using a GA. The crossover is done in the usual way. While in the mutation step of the GA, the information obtained from the collected data is used to guide the search algorithm to focus on factors, and levels of factors that improve the fitness criterion for the search. In the weighted mutation scheme of the SELC algorithm, significant main effects and pairwise interactions are identified. If a factor, $F_j$, has a significant main effect and no significant pairwise interactions, then the mutation probability ($p_{jl}$) for each level, $l$, of the factor is proportional to the average fitness of that level for the data collected thus far in the experiment:

$$p_{j_l} \propto \bar{y}(F_j = l) \qquad \text{for } j = 1, 2, \ldots, J, \text{ and } l = 1, 2, \ldots, L.$$

If the two factors, $F_j$ and $F_k$, have a significant interaction, and either of the two factors is chosen, the mutation is weighted jointly with probability

$$p_{j_l k_m} \propto \bar{y}(F_j = l, F_k = m) \qquad \text{for } j, k = 1, 2, \ldots, J, \text{ and } l, m = 1, 2, \ldots, L.$$

If the selected factor does not have a significant main effect or interaction, then its value will be changed to any possible level with equal probability.

In the SELC algorithm, the concept of forbidden arrays and weighted mutations are combined with standard genetic algorithms. The algorithm starts by selecting an initial design based on an appropriate orthogonal array. If there is prior knowledge about the design space or design points, the compounds with low process values are included in the forbidden array. The desired compounds in the initial design are then manufactured and the forbidden array is updated. New offspring (compounds) are selected using a genetic algorithm with weighted mutation probabilities. Similar to many optimization algorithms, the stopping rule for the SELC algorithm is also subjective and depends on the progression of the algorithm and experimental constraints. Next, we illustrate the limitations of SELC in terms of identifying good compounds.

*4D example (four dimensional problem).* Consider a pharmaceutical experiment setup, similar to that in Section 2, where the core molecule has four locations $A$, $B$, $C$ and $D$, and ten monomers can be added to each of the four locations. This corresponds to an experiment with four factors each at ten levels $1 \leq x_{\cdot j} \leq 10$, $j = 1, \ldots, 4$. Further suppose that the response of interest $y(x_i)$ is a scalar valued physiochemical property of a compound, where $x_i = (x_{i1}, x_{i2}, x_{i3}, x_{i4})$. For illustration purposes, we generate the responses using the 4-dimensional Levy function (see Section 5 for details). Since the true process values are known for all the $10^4$ possible compounds (the entire compound library), the performance of SELC for finding the maximum process value can be evaluated via simulation. We used 500 simulations on designs of run size 150 each. For each simulation, a 40-run minimax design [Johnson et al. (1990); John et al. (1995)] was randomly chosen for an initial design. We used SELC to obtain the rest of the 110 designs points in batches of size $b = 6$, with the exception of the last batch of two trials only. On average, SELC successfully identifies the true global maximum only 23.4% of the times. In many situations (including the pharmaceutical chemistry example), interest lies in obtaining a few good candidates instead of the "absolute best." The success rate for capturing the top five maxima (including the global maximum) is 83.0%. As we shall see in Section 5, $\mathcal{G}$-SELC demonstrates a much better performance.

3.2. *Expected improvement approach.* In the innovative approach developed in Jones et al. (1998), a stochastic process (Gaussian spatial process) is used to model the underlying process. They developed a sequential design strategy based on a figure of merit called the *expected improvement function* [first introduced in Mockus, Tiesis and Zilinskas (1978)]. This technique often requires the fewest function evaluations compared to several competing methods. Next, we describe two major components of this procedure: (a) the Gaussian process model used to get a surrogate of the underlying process and (b) the expected improvement function.

Although the modeling in Jones et al. (1998) was developed for a continuous hyper-rectangle, the same modeling approach can be applied for both discretized and continuous convex design regions. The $n \times d$ experiment design matrix, $X$, is the matrix of input trials, where the $i$th input trial is a $d$-dimensional vector $x_i = (x_{i1}, \ldots, x_{id})$. The outputs for the simulation trials $y = y(X) = (y_1, y_2, \ldots, y_n)'$ are modeled as

$$(1) \qquad y(x_i) = \mu + z(x_i); \qquad i = 1, \ldots, n,$$

where $\mu$ is the overall mean, and $z(x_i)$ is a Gaussian spatial process with $E(z(x_i)) = 0$, $\mathrm{Var}(z(x_i)) = \sigma_z^2$, and $\mathrm{cov}(z(x_i), z(x_j)) = \sigma_z^2 R_{ij}$. Jones et al. (1998) used power exponential correlation given by

$$(2) \qquad \begin{aligned} R_{ij} &= \mathrm{corr}(z(x_i), z(x_j)) \\ &= \prod_{k=1}^{d} \exp\{-\theta_k(x_{ik} - x_{jk})^{p_k}\} \qquad \text{for all } i, j, \end{aligned}$$

to model the correlation structure of the underlying process, where the exponent, $p_k$, is the smoothness parameter in the direction of the $k$th factor, $k = 1, \ldots, d$, and $\theta = (\theta_1, \ldots, \theta_d)$ is the vector of hyper-parameters. The power exponential correlation structure, for different values of $p_k$ and $\theta$, generates a large class of correlation functions. Another commonly used correlation structure known as the product Matérn correlation function [Stein (1999); Santner, Williams and Notz (2003)] can be used instead. It turns out that, for the application in Section 2 and the examples in Section 5, both the product Matérn correlation and the power exponential family behave very similarly. To save space, here we use the power exponential correlation function. Specifying $p_k = 2$ in the power exponential correlation (also known as the Gaussian correlation) is a reasonable simplifying assumption for the applications in this paper. Furthermore, if one or more factors (or input variables) are qualitative, which is not very rare for pharmaceutical experiments, one can use the correlation function proposed by Qian, Wu and Wu (2008) to model the correlation structure.

In general, $y(X)$ has multivariate normal distribution, $y(X) \sim N_n(\mathbf{1_n}\mu, \Sigma)$, where $\Sigma = \sigma_z^2 R$ and $R = [R_{ij}]$. The Gaussian process model can be used to estimate responses at any nonsampled point $x^*$. The best linear unbiased predictor (BLUP) for $y(x^*)$ is

$$\hat{y}(x^*) = \hat{\mu} + r'R^{-1}(y - \mathbf{1_n}\hat{\mu}), \tag{3}$$

[see Santner, Williams and Notz (2003) for details] with mean squared error

$$s^2(x^*) = \sigma_z^2 \left(1 - r'R^{-1}r + \frac{(1 - \mathbf{1'_n}R^{-1}r)^2}{\mathbf{1'_n}R^{-1}\mathbf{1_n}}\right), \tag{4}$$

where $r = (r_1(x^*), \ldots, r_n(x^*))'$, and $r_i(x^*) = \text{corr}(z(x^*), z(x_i))$ is defined in equation (2). In practice the parameters are replaced with the maximum likelihood estimates. For details on parameter estimation, uncertainty in prediction and other model properties, see Jones et al. (1998) and Ranjan, Bingham and Michailidis (2008).

Jones et al. (1998) argue that finding a global optimum of a process $y(x)$ by simply using the optima of its BLUP is not a good idea, as it does not acknowledge the model uncertainty. They propose using a figure of merit called "expected improvement" (EI), which balances local and global search. Let $f_{\max}$ be the current estimate of the global maximum. Then, as in Jones et al. (1998), the improvement in the estimate of the process maximum, by including the design point $x$ in the current sample, can be written as

$$I(x) = \max\{y(x) - f_{\max}, 0\}. \tag{5}$$

In our context, the improvement in the estimate of the highest process value, by manufacturing a new compound, can be obtained using equation (5). Jones et al. (1998) argue that by taking the expectation of the improvement function, uncertainty in the model is taken into account. This formulation ensures the exploration of the design space both inside and outside the neighborhood of the current maximum $f_{\max}$. The corresponding expected improvement at $x \in \chi$ (the design space) is given by

$$E[I(x)] = s^2(x)\phi\left(\frac{\hat{y}(x) - f_{\max}}{s(x)}\right) + (\hat{y}(x) - f_{\max})\Phi\left(\frac{\hat{y}(x) - f_{\max}}{s(x)}\right), \tag{6}$$

where $\phi(\cdot)$ and $\Phi(\cdot)$ are the standard normal probability density function and cumulative distribution function respectively [see Jones et al. (1998) and Ranjan et al. (2008) for details].

Since the proposed technique in Jones et al. (1998) is a complete sequential strategy (one trial at a time), and developing an efficient EI-based batch sequential design is in itself a challenging problem, we adapt the existing procedure to select a batch of $b$ trials. This is done by first choosing $\lceil \alpha b \rceil$ trials ($0 \leq \alpha \leq 1$) from the top, instead of the top most design point, in the candidate set ranked according to their EI values. Here, $\lceil w \rceil$ denotes the smallest integer greater than or equal to $w$. Note that selecting more than one trial from the top of the sorted EI vector will not be beneficial if the EI function is either unimodal, or its highest peak is much higher than the other peaks. In the illustrative examples and the real applications considered in this paper, EI functions are often multimodal with comparable heights. The following example illustrates the performance of this methodology for the pharmaceutical scenario presented in Section 3.1.

*4D example (contd.).* Instead of using SELC, we use the adapted EI approach for choosing the additional 110 trials in batches of size 6 each. Similar to the illustration of SELC, the average performance of this algorithm was observed based on 500 simulations. The adapted EI algorithm successfully identifies the true global maximum only 19.6% of the times, and the overall success rate for capturing the top five maxima (including the global maximum) is 24.6%.

Clearly, the ability of this algorithm for capturing the global maxima, along with some near-optima, is not outstanding. This is not surprising because the improvement function defined in equation (5) is targeted for the global maximum. Earlier we have seen that the performance of SELC alone is not outstanding either. Now we propose a new approach that improves upon the SELC algorithm using the EI-based sampling scheme. The SELC part of it captures the good compounds that may not be the best one, and the adapted EI part targets the best compound and minimizes the overall model uncertainty.

**4. $\mathcal{G}$-SELC: New algorithm—GP based SELC.** A new batch-sequential methodology for constructing a subset of the compound library is now developed. In short, the new algorithm uses an adaptive mixture of the batches of trials obtained from the two approaches, where the mixing ratio $\alpha$ depends on the peakedness and modality of the current estimate of the predicted surface. We call this new algorithm $\mathcal{G}$-SELC.

Suppose the compound library consists of $M$ compounds that can be manufactured if required. Assuming that the process evaluation (i.e., creation of a new compound) is expensive, the budget on the number of compounds $N (\ll M)$ to be created is often fixed. Further assume that it is preferred to manufacture compounds in batches of size $b$ each. Considering these constraints, we construct a subset of the combinational library by first selecting an initial design of size $n = n_0$ to get an overall idea of the underlying process. We use a minimax design [Johnson et al. (1990); John et al. (1995)] to construct an initial design of size $n_0$. Alternatively, one can use other space-filling designs (e.g., orthogonal array based Latin Hypercube designs [Tang (1993)] and uniform coverage designs [Lam et al. (2002)]). The data obtained from the initial set of compounds is then used to fit a Gaussian process model to estimate the underlying process. Next, the peakedness and the modality of the surrogate is used to estimate the mixing ratio $\alpha$.

If $b$ compounds are selected at each stage, $\lceil \alpha b \rceil$ of them are selected using the adapted EI criterion and the remaining $b - \lceil \alpha b \rceil$ compounds by SELC. Note that the EI-part of the $\mathcal{G}$-SELC performs a directed search, whereas the SELC-part does it randomly and explores the unexplored regions of the design space more frequently. If the fitted response surface is relatively simple and has only one peak, that is, there is only one cluster of good compounds, then it is logical to choose one (or, very few) candidate(s) in the region of best predicted value based on EI-part, and utilize the remaining candidates to explore other regions using the SELC-part. Similarly, if there are two distinguishable peaks of the predicted response surface (i.e., there are two distinct clusters of points with high predicted value), then it is natural to select twice as many candidates compared to the previous case, and to choose the remaining runs randomly (using SELC) from the unexplored regions.

This motivates the formulation of the mixing ratio $\alpha$. Although the application at hand has discretized space, we develop the theory of mixing ratio for a more general setup. Let $\chi$ be a convex search space. Define

$$(7) \qquad S = \{x \in \chi : \hat{y}(x) > c f_{\max}\}$$

to be the region that consists of design points with high process values, where $c < 1$ and $f_{\max} > 0$ is the current estimate of the process maximum. Of course, the choice of $c$ is subjective and can be based on prior information about the underlying process. Let $k$ be the number of distinguishable clusters of points in $S$. Note that it is possible that the underlying process has more than $k$ distinct modes, and the undetected peaks are not high enough to include design points with high process values. We use a $k$-means clustering algorithm to identify the clusters and cross-validation techniques to minimize the total probability of misclassification in order to find the right number of clusters. If $C_i$'s are the clusters such that $S = \bigcup_{i=1}^{k} C_i$ and $C_i \cap C_j = \phi$ for all $i \neq j$, then the mixing ratio $\alpha$ can be defined as

$$\alpha = \sum_{i=1}^{k} \frac{Ar(C_i)}{Ar(\chi)},$$

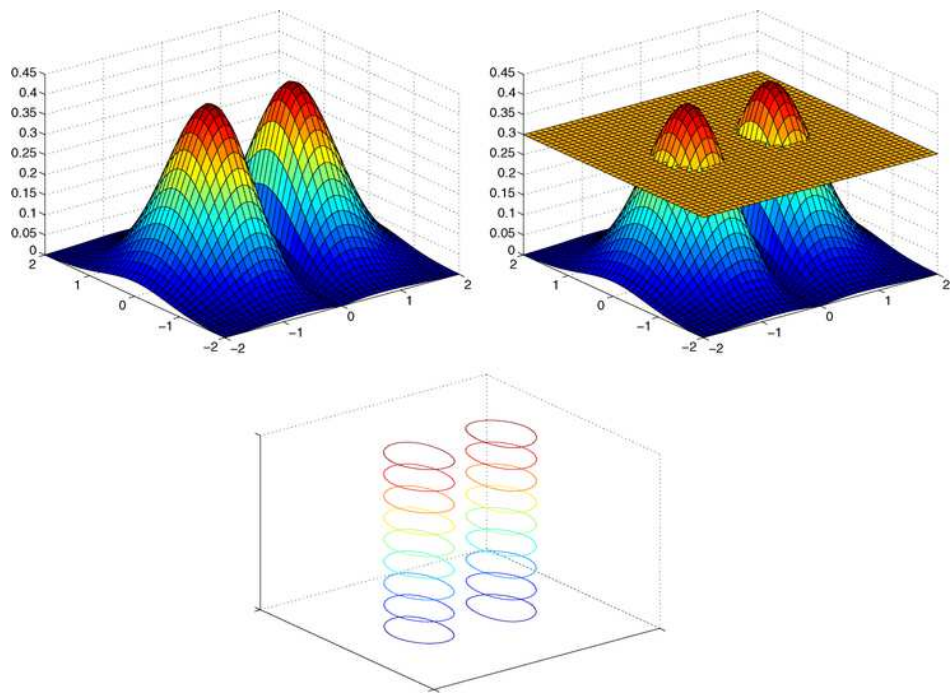

Fig. 2. *The fitted response function has two peaks—at height 0.3 there are two clear clusters as demonstrated on the right most panel.*

where $Ar(R)$ is the area/volume of the region $R$. It is easy to verify that $0 < \alpha < 1$. If the design space is discrete (e.g., a fine grid, or a compound library in our case), it is much easier to compute the value of $\alpha$. Using Monte Carlo approximation of the $Ar(C_i)$'s, one can simply estimate $\alpha$ by $\hat{\alpha} = \frac{|S|}{|\chi|}$, where $|R|$ is the cardinality of the set $R$. The batch of $b$ new compounds to be manufactured is a mixture of $\lceil \alpha b \rceil$ compounds from the EI-based sampling scheme described above, and $b - \lceil \alpha b \rceil$ compounds are selected using the SELC technique.

Let us illustrate this with a hypothetical example as depicted in Figure 2. Let $\chi = [-2, 2]^2$ be the design region. The left most panel illustrates that the fitted surface has two distinct peaks of comparable heights, and the figures in the other two panels show that the set $S$ has $k = 2$ distinguishable clusters. Under these settings, $Ar(\chi) = 4 \times 4 = 16$ and $Ar(C_1) = Ar(C_2) \simeq \pi(\frac{1}{2})^2 = 0.785$. Thus,

$$\alpha = \frac{0.785 + 0.785}{16} = \frac{1.57}{16} = 0.098,$$

and $\lceil \alpha b \rceil = 2$ if $b = 16$. That is, in a batch of 16 new compounds, 2 new compounds will be found using the adapted EI method and the remaining 14 compounds using the SELC algorithm. The batch-sequential sampling mechanism using $\mathcal{G}$-SELC is summarized as follows:

1. Choose an initial design of size $n = n_0$, such that $n_0$ is a small fraction of the maximum allowable budget $N$. Usually, $N$ is much smaller than $M$, the total number of candidate design points (or the size of compound library, in our context).
2. Fit a surrogate using Gaussian process model (1)–(4) on $n$ data points. Let $\hat{y}(x)$, $x \in \chi$ be the predicted surface.
3. Construct the region $S$ and the clusters $C_i$, $i = 1, \ldots, k$. Then, evaluate the mixing ratio $\alpha$.
4. Choose $\lceil \alpha b \rceil$ best trials by sorting the set $\{E[I(x_i)], i = 1, \ldots, M-n\}$ and $b - \lceil \alpha b \rceil$ trials using the SELC algorithm to construct the batch of $b$ new compounds $x_{\text{new}} = \{x_{i,\text{new}}, i = 1, \ldots, b\}$.
5. Manufacture the compounds corresponding to the new trials $x_{\text{new}}$. Compute the corresponding responses $y_{\text{new}} = \{y(x_{i,\text{new}}), i = 1, \ldots, b\}$.
6. Update the data $x = [x' : x'_{\text{new}}]'$, $y = [y' : y'_{\text{new}}]'$ and $n = n + b$.
7. Repeat Step 2 to Step 6, until the budget allows or the optimum is achieved with desired level of accuracy.

Note that the mixing ratio $\alpha$ changes from iteration to iteration and stabilizes as the number of iterations increases. Next, we illustrate the findings through some simulated examples and the pharmaceutical chemistry example.

**5. Simulated examples.** To illustrate the relative performance of the proposed approach compared to SELC and adapted EI, several examples are now presented. We use random minimax designs (any good space filling design can be used instead) with $n_0 = 10 \times d$ trials for initial designs. The results reported in this section use the Gaussian correlation function for GP modeling. The mixing ratio $\alpha$ is computed by taking $c = 3/4$ in equation (7). Note that the choice of a batch size, $b$, may impact the performance of the proposed approach. In each of the examples, several choices of $b$ are considered and the performance is observed. In practice, the batch size may be fixed beforehand because of the restrictions imposed by the experimenter. We also present a few plots which aid in the evaluation of the approaches.

*4D example (contd.).* The performance of all the three search algorithms is now assessed on their ability to find the global maximum and the other top four maxima where the responses are generated using the 4-dimensional Levy function [Levy and Montalvo (1985)]:

$$y(x_1, \ldots, x_d) = \sin^2\left\{\pi\left(\frac{x_1 + 2}{4}\right)\right\}$$

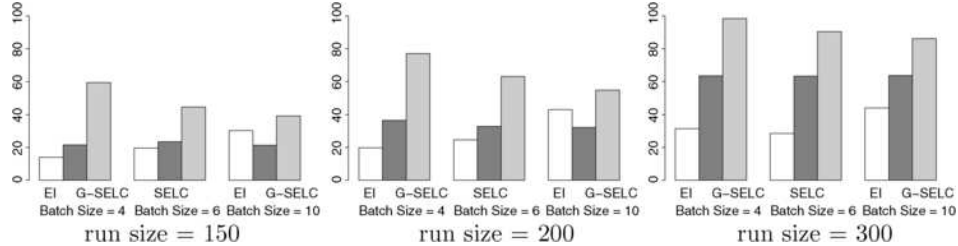

FIG. 3. *Performance of the three algorithms—percentage of success in identifying maxima: in each block, the left, middle and right columns represent EI, SELC and $\mathcal{G}$-SELC respectively.*

$$+ \sum_{i=1}^{d-1} \left(\frac{x_i - 2}{4}\right)^2 \left\{1 + 10\sin^2\left(\pi\left(\frac{x_i + 2}{4}\right) + 1\right)\right\}$$
$$+ \left(\frac{x_d - 2}{4}\right)^2 \left\{1 + \sin^2(2\pi(x_d - 1))\right\}.$$

The Levy function is commonly used as a test function in the global optimization literature [Boggs (1985)]. The performance of all the three algorithms is averaged over 500 simulations with different run sizes and batch sizes. The results are summarized in Table 4. The last two columns of Table 4 contain the success rate of capturing the global maximum and the overall performance in identifying the top 5 maxima (including the global maximum).

It is clear from the last two columns of Table 4 that $\mathcal{G}$-SELC outperforms both SELC and the adapted version of EI. For example, when the total run size is 150 and batch size 4, $\mathcal{G}$-SELC finds the top five candidates 93.6% of times, whereas SELC finds them 81.4% of the times and the success rate for EI is less than 15%. The last column "Total" shows that the relative performance of $\mathcal{G}$-SELC over SELC and EI improves as the batch size $b$ decreases. As one might guess, the performance of all the three algorithms increases with the increase in run size. Interestingly, the performance of the adapted EI improves with the increase in batch size. This is somewhat counterintuitive, as EI is expected to perform well for small batch sizes, because in that case the surface estimate is updated more frequently.

Figure 3 depicts the success rates of the three algorithms for achieving the maximum for different run sizes and batch sizes as shown in the "Max" column of Table 4. The left most panel presents the results when the search is stopped after 150 runs, the middle and the right most panel correspond to run size 200 and 300, respectively. In each of the three panels, the three blocks (from left to right) represent batch sizes 4, 6 and 10, respectively. The left most column in each block (i.e., with a fixed run size and batch size) represents the performance of EI, the middle one corresponds to SELC and the right most one represents 𝒢-SELC.

TABLE 4
*Performance of the three different methods based on 500 simulations—percentage success in identifying the top 5 maxima of the Lévy function evaluated on a grid of $10^4$ points*

| Run size | Batch size $b$ | Method | 5th best | 4th best | 3rd best | 2nd best | Max | Total |
|---|---|---|---|---|---|---|---|---|
| 150 | 4 | EI | 0.0 | 0.0 | 0.0 | 0.4 | 14.0 | 14.4 |
| | | SELC | 14.6 | 1.4 | 33.0 | 10.8 | 21.6 | 81.4 |
| | | 𝒢-SELC | 4.2 | 0.2 | 21.8 | 7.8 | 59.6 | 93.6 |
| | 6 | EI | 0.2 | 0.0 | 3.4 | 1.4 | 19.6 | 24.6 |
| | | SELC | 12.0 | 1.6 | 31.6 | 14.4 | 23.4 | 83.0 |
| | | 𝒢-SELC | 5.0 | 0.2 | 25.0 | 15.0 | 44.6 | 89.8 |
| | 10 | EI | 0.4 | 0.0 | 10.0 | 0.8 | 30.2 | 41.4 |
| | | SELC | 11.6 | 1.0 | 33.6 | 14.0 | 21.4 | 81.6 |
| | | 𝒢-SELC | 5.2 | 1.2 | 30.2 | 15.8 | 39.2 | 91.6 |
| 200 | 4 | EI | 0.0 | 0.0 | 0.0 | 0.0 | 19.8 | 19.8 |
| | | SELC | 6.4 | 2.2 | 27.6 | 20.6 | 36.4 | 93.2 |
| | | 𝒢-SELC | 1.6 | 0.2 | 10.8 | 8.2 | 77.2 | 98.0 |
| | 6 | EI | 0.0 | 0.0 | 0.0 | 0.2 | 24.6 | 24.8 |
| | | SELC | 8.0 | 2.0 | 27.4 | 20.2 | 32.8 | 90.4 |
| | | 𝒢-SELC | 1.6 | 0.2 | 18.4 | 13.4 | 63.2 | 96.8 |
| | 10 | EI | 0.0 | 0.0 | 0.0 | 0.0 | 43.0 | 43.0 |
| | | SELC | 10.6 | 0.4 | 30.8 | 18.2 | 32.2 | 92.2 |
| | | 𝒢-SELC | 3.6 | 0.6 | 21.4 | 18.8 | 54.8 | 99.2 |
| 300 | 4 | EI | 2.5 | 0.0 | 3.5 | 1.0 | 31.5 | 38.5 |
| | | SELC | 3.0 | 0.0 | 11.0 | 22.0 | 63.5 | 99.5 |
| | | 𝒢-SELC | 0.0 | 0.0 | 1.5 | 0.0 | 98.5 | 100.0 |
| | 6 | EI | 0.0 | 0.0 | 0.5 | 0.5 | 28.6 | 29.5 |
| | | SELC | 1.4 | 1.0 | 14.3 | 18.1 | 63.3 | 98.1 |
| | | 𝒢-SELC | 0.0 | 0.0 | 3.8 | 5.7 | 90.5 | 100.0 |
| | 10 | EI | 0.0 | 0.6 | 0.4 | 0.8 | 44.0 | 45.8 |
| | | SELC | 2.6 | 0.4 | 11.8 | 20.6 | 63.8 | 99.2 |
| | | 𝒢-SELC | 0.6 | 0.0 | 4.2 | 9.0 | 86.2 | 100.0 |

It should be noted that, for each run size, the performance of SELC remains nearly constant (which indicates insignificant "learning") as the batch size increases. The performance of 𝒢-SELC, for finding the maximum, decreases with the increase in batch size, whereas the trend is reversed for EI. Furthermore, the adapted EI is quite powerful in identifying the global maximum and not so good in finding other near-optima, which is not surprising.

TABLE 5
*Performance of the three different methods based on 100 simulations—percentage of success in identifying maxima of the Pavini function evaluated on a grid of $10^5$ points*

| Order of forbidden array | Run size | Method | 25.59 (3rd best) | 25.60 (2nd best) | 25.62 (Max) | Total |
|---|---|---|---|---|---|---|
| 3 | 150 | EI | 4 | 0 | 0 | 4 |
|   |     | SELC | 0 | 1 | 0 | 1 |
|   |     | $\mathcal{G}$-SELC | 7 | 2 | 0 | 9 |
|   | 200 | EI | 19 | 3 | 1 | 23 |
|   |     | SELC | 1 | 0 | 0 | 1 |
|   |     | $\mathcal{G}$-SELC | 20 | 9 | 0 | 29 |
| 4 | 150 | EI | 4 | 0 | 0 | 4 |
|   |     | SELC | 0 | 1 | 0 | 1 |
|   |     | $\mathcal{G}$-SELC | 4 | 0 | 0 | 4 |
|   | 200 | EI | 20 | 4 | 0 | 24 |
|   |     | SELC | 2 | 0 | 1 | 3 |
|   |     | $\mathcal{G}$-SELC | 24 | 6 | 0 | 30 |

*5D example (five dimensional problem).* Similar to the 4D example setup, consider a pharmaceutical experiment where a core molecule has five locations $A$, $B$, $C$, $D$ and $E$, and ten monomers can be added to each of the five locations. That is, $1 \leq x_{\cdot j} \leq 10$ for $j = 1, \ldots, 5$, and the response of interest $y(x_i)$ is the desired chemical property of a compound with monomers specified by $x_i = (x_{i1}, x_{i2}, x_{i3}, x_{i4}, x_{i5})$. Now suppose that the responses were generated using the Paviani function [Andersen, Nielsen and Kreiborg (1998)]

$$f(x) = \sum_{i=1}^{5}(\ln^2(x_i) + \ln^2(11 - x_i)) - \left(\prod_{i=1}^{5} x_i\right)^{0.2}.$$

The performance of the three approaches is assessed based on its ability to achieve the top 3 maxima of $f(x)$, given by 25.59, 25.60 and 25.62. The implementation starts with first finding a random minimax design of size $10 \times d = 50$ for the initial design. The rest of the compounds are constructed in batches of 10 until the budget allows. For this example, we consider both forbidden arrays of order 3 and 4. The success rate of the three approaches (SELC, EI and $\mathcal{G}$-SELC) is averaged over 100 simulations.

None of these methods perform well for run size 150. This is not surprising, considering the fact that here we are evaluating only 0.15% of all eligible candidates. It is also expected that SELC will not perform well here because the evolutionary algorithms would require more runs to reap the benefits. As the top three maxima have very similar values, the last column of Table [5](5) is more informative. We see that the proposed method outperforms SELC and adapted EI.

This example illustrates the limitations of SELC on a bigger search space (i.e., the ratio $N/M$ defined in Section [4](4) is really small). SELC does not perform well simply because it cannot evaluate enough candidates. Under such circumstances, the power of 𝒢-SELC lies mostly in its "systematic part" which borrows strength from EI.

**6. 3D example (pharmaceutical chemistry example) revisited.** In this section we illustrate the performance of 𝒢-SELC for the pharmaceutical chemistry example discussed in Section [2](2). The goal is to find a subset of the compound library that contains good compounds and hopefully the compound(s) with highest process value. Recall that the compound library consists of a total of 40,970 ($= 5 \times 34 \times 241$) compounds. However, the response, $y$, values are available for only 1800 of these compounds [see Mandal et al. (2007) for details]. Under these circumstances, the implementation of any methodology is restricted to the exploration of only this set of 1800 compounds. Ideally, one should suggest a subset of good compounds using the information on the entire compound library.

The procedure starts by first finding a 50-point random minimax design (Table [6](6)) for creating an initial set of compounds that provides an overall idea of the underlying process. Once the data for an initial design is obtained, the Gaussian process model described in Section [3.2](3.2) is used to get a surrogate of the underlying process. If the factors (input variables) are qualitative with $k$ levels, we treat them as quantitative with values $1, 2, \ldots, k$. Then, we use 𝒢-SELC to choose 48 more compounds in 12 batches of size 4 each. The mixing ratio $\alpha$ is computed by taking $c = 3/4$ in equation ([7](7)). These 48 compounds from the follow-up runs along with the 50 compounds from the initial design constitute the required subset of the compound library.

Based on the prior knowledge of the scientists, compounds with monomer structure shown in Table [7](7) are also known to be uninteresting, and are therefore placed in the forbidden array.

The mutation probabilities are weighted according to the average substructure performance of the initial design (see Table [8](8)). Each substructure (monomer) at location A receives the same baseline weighted mutation probability ($0.25 \times \frac{1}{4}$). For the substructures with positive average response, additional weights are assigned. For example, the additional weight for substructure 2 is ($0.8 \ / \ (0.8 + 0.36) = 0.69$). The monomers for locations $B$ and $C$ are treated similarly.

Table [9](9) presents the 48 new compounds obtained using 𝒢-SELC with the afore-described forbidden array and mutation scheme. The left panel in Table [9](9) shows the 24 compounds obtained from the EI steps, and the right panel presents the compounds suggested by the SELC steps.

TABLE 6
*Initial design: set of compounds selected to fit a surrogate of the underlying process*

| A | B | C | y | A | B | C | y |
|---|---|---|---|---|---|---|---|
| 2 | 20 | 35 | 4 | 3 | 2 | 25 | 5 |
| 2 | 15 | 39 | 2 | 3 | 18 | 22 | −3 |
| 2 | 15 | 9 | 5 | 3 | 9 | 19 | 0 |
| 2 | 20 | 41 | −1 | 3 | 12 | 1 | −4 |
| 2 | 2 | 15 | −3 | 3 | 21 | 25 | 1 |
| 2 | 20 | 1 | 0 | 3 | 23 | 19 | −1 |
| 2 | 23 | 6 | 4 | 3 | 3 | 7 | −2 |
| 2 | 23 | 12 | 2 | 3 | 3 | 35 | 0 |
| 2 | 3 | 20 | −3 | 3 | 8 | 10 | 0 |
| 2 | 14 | 21 | −2 | 3 | 13 | 11 | 0 |
| 3 | 16 | 27 | −9 | 3 | 19 | 30 | 17 |
| 3 | 9 | 39 | 33 | 3 | 14 | 16 | 1 |
| 3 | 6 | 39 | 1 | 3 | 8 | 2 | 3 |
| 3 | 7 | 30 | −5 | 3 | 12 | 31 | 0 |
| 3 | 23 | 31 | 0 | 3 | 21 | 38 | 10 |
| 3 | 8 | 6 | 1 | 3 | 21 | 17 | −7 |
| 3 | 18 | 13 | 7 | 3 | 3 | 11 | −8 |
| 3 | 8 | 28 | −3 | 3 | 18 | 4 | −7 |
| 3 | 2 | 39 | −3 | 3 | 9 | 14 | −1 |
| 3 | 10 | 25 | −3 | 3 | 12 | 38 | −2 |
| 3 | 8 | 34 | 4 | 3 | 3 | 2 | −4 |
| 3 | 18 | 10 | 3 | 3 | 7 | 22 | 0 |
| 3 | 13 | 6 | 2 | 3 | 2 | 31 | −6 |
| 3 | 6 | 15 | −1 | 3 | 15 | 33 | −3 |
| 3 | 5 | 26 | −2 | 4 | 13 | 38 | −10 |

TABLE 7
*Forbidden array from prior knowledge, for the pharmaceutical example*

| A | B | C | A | B | C |
|---|---|---|---|---|---|
| 1 | 19 | 10 | 3 | 21 | 22 |
| 1 | 19 | 22 | 3 | 21 | 23 |
| 1 | 21 | 10 | 4 | 19 | 10 |
| 3 | 10 | 3 | 4 | 19 | 20 |
| 3 | 19 | 10 | 4 | 19 | 28 |
| 3 | 19 | 22 | 4 | 21 | 20 |
| 3 | 21 | 20 | 4 | 21 | 23 |

Note that among these 48 new compounds, 3 compounds are good, that is, with response $y$ greater than 10. The best compound with $y = 48$ ($A = 4, B = 21$,
$C = 30$) has also been identified in these iterations. Of course, if the search

space was not limited to only 1800 compounds, more good compounds are likely to be captured in the batches of compounds. To compare the performance of the proposed approach with SELC and the adapted EI algorithm, we repeated this procedure 100 times. That is, starting with the same initial design of size 50, we added 48 compounds sequentially in batches of size 4 each proposed by the $\mathcal{G}$-SELC algorithm. Then, the number of good compounds in the follow-up trials are recorded. Similarly, the number of good compounds were recorded when SELC and adapted EI were used instead of $\mathcal{G}$-SELC. Results are summarized in Table 10.

In terms of capturing the maxima, it is clear that all the three approaches perform poorly for this data set. This is not very surprising as the search space is limited to only 1800 compounds and we are exploring only 5.5% of them. Note that the adapted EI algorithm performs better for this application in identifying the global maximum. Since identifying "good" compounds is relatively more important compared to finding the "single best" compound in this application, the performance of these algorithms can be assessed based on the number of "good" compounds identified. The last row of Table 10 represents the total number of good compounds (compounds with response $y$ greater than 10) obtained in 100 simulations. It is clear that $\mathcal{G}$-SELC identifies almost twice as many good compounds compared to the other two techniques.

**7. Summary and concluding remarks.** The identification of promising compounds from a large compound library is often very expensive. The proposed approach $\mathcal{G}$-SELC does this job efficiently, and requires exploration of much fewer compounds from the library compared to other competing methods. Of course, the relative performance of $\mathcal{G}$-SELC depends on the complexity of the underlying process. If the response surface is very smooth, any reasonable search algorithm should work satisfactorily. For an extremely complicated surface, almost complete enumeration might be needed irrespective of the efficiency of the search methods. For response surfaces whose ruggedness lies in between, $\mathcal{G}$-SELC is expected to perform well.

TABLE 8
*Weighted mutation probabilities for each substructure at Position A, for the pharmaceutical example*

| Substructure | Average response | Weighted mutation probability |
|---|---|---|
| 1 | NA | $= 0.25 \times \frac{1}{4} + 0 \times \frac{3}{4}$ |
| 2 | 0.80 | $= 0.25 \times \frac{1}{4} + 0.69 \times \frac{3}{4}$ |
| 3 | 0.36 | $= 0.25 \times \frac{1}{4} + 0.31 \times \frac{3}{4}$ |
| 4 | $-10.00$ | $= 0.25 \times \frac{1}{4} + 0 \times \frac{3}{4}$ |

TABLE 9
*Suggested new compounds for the pharmaceutical chemistry example*

| $A$ | $B$ | $C$ | $y$ | $A$ | $B$ | $C$ | $y$ |
|---|---|---|---|---|---|---|---|
| 1 | 19 | 29 | $-2$ | 1 | 1 | 10 | $-3$ |
| 1 | 21 | 30 | 4 | 1 | 1 | 35 | 6 |
| 2 | 9 | 39 | 0 | 1 | 1 | 39 | 4 |
| 2 | 20 | 30 | 5 | 1 | 7 | 2 | $-5$ |
| 3 | 8 | 39 | 4 | 1 | 7 | 21 | $-9$ |
| 3 | 9 | 38 | $-4$ | 1 | 10 | 39 | 6 |
| 3 | 9 | 40 | $-2$ | 1 | 18 | 8 | $-3$ |
| 3 | 10 | 38 | $-8$ | 1 | 19 | 4 | 6 |
| 3 | 10 | 39 | 24 | 1 | 19 | 6 | 4 |
| 3 | 10 | 40 | 3 | 1 | 19 | 8 | 5 |
| 3 | 18 | 30 | $-3$ | 1 | 19 | 13 | 4 |
| 3 | 19 | 29 | 3 | 1 | 19 | 21 | 0 |
| 3 | 19 | 31 | $-1$ | 1 | 19 | 30 | 9 |
| 3 | 19 | 39 | 0 | 1 | 19 | 39 | 22 |
| 3 | 20 | 30 | $-7$ | 1 | 19 | 40 | 2 |
| 3 | 20 | 38 | $-4$ | 2 | 2 | 25 | $-4$ |
| 3 | 21 | 30 | 9 | 2 | 9 | 8 | $-2$ |
| 4 | 9 | 39 | 1 | 2 | 9 | 39 | 0 |
| 4 | 10 | 39 | 0 | 2 | 15 | 10 | $-1$ |
| 4 | 20 | 30 | $-18$ | 2 | 23 | 39 | $-4$ |
| 4 | 21 | 29 | $-7$ | 3 | 8 | 1 | 1 |
| 4 | 21 | 30 | 48 | 3 | 10 | 6 | 5 |
| 4 | 21 | 31 | 1 | 3 | 18 | 14 | 1 |
| 4 | 23 | 30 | $-3$ | 3 | 21 | 5 | $-4$ |

The implementation of $\mathcal{G}$-SELC starts with finding a set of compounds using a good $n_0$-point space-filling design (we used random minimax designs), where $n_0$ is a small fraction of the total budget ($N$) on the number of compounds to be manufactured. The modeling (1)–(4) of the data, obtained from this set of compounds, provides an overall idea of the underlying chemical process of interest. This is followed by finding the rest of the compounds in batches of $b$ compounds. $\mathcal{G}$-SELC suggests that a batch with an adaptive mixture of compounds proposed by SELC and adapted-EI is likely to contain more "good" compounds compared to either of the two techniques on their own. In the motivating pharmaceutical example, we observed that $\mathcal{G}$-SELC identifies almost twice as many "good" compounds compared to the other two techniques.

TABLE 10
*Simulation results for the pharmaceutical example: Success rates of different methods for identifying good compounds*

| | EI | SELC | $\mathcal{G}$-SELC |
|---|---|---|---|
| | Success rate for identifying good compounds | | |
| Third best | 92 | 100 | 95 |
| Second best | 0 | 0 | 4 |
| Best | 8 | 0 | 1 |
| | Total number of good compounds in 100 simulation | | |
| | 166 | 154 | 314 |

A few additional remarks are worth noting. First, the EI formulation used here is geared toward finding the best, and not other near-optimal solutions. Thus, if properly formulated, the EI-based approach is likely to perform better than the one used here, and will boost the performance of $\mathcal{G}$-SELC. Since formulating a new EI technique was not the motive here, we used an adapted version of the existing approach.

Second, the Gaussian process model used here does not assume any prior information regarding the behavior of the molecules. Since no such information were available to us for the pharmaceutical example discussed in Section 2, we used a constant $\mu$. If scientists have prior information on the shape of the underlying process, one could use a deterministic function $\mu(x)$ instead of a constant $\mu$. Such models are also known as blind kriging [Joseph, Hung and Sujiantao (2008)].

Although we have used the power exponential correlation function (Gaussian correlation function) throughout the paper, we investigated the performance of $\mathcal{G}$-SELC for the product Matérn correlation function with different values of the smoothness parameter $\nu = 2.5$, 3.5 and 5.5. It turns out that, for the motivating pharmaceutical example in Section 6 and the simulated examples in Section 5, the two correlation structures lead to very similar results for all the three methodologies presented here. Further investigations on the distribution of the standardized predictions errors show that the two types of correlation functions result in very similar predictions for several test functions (e.g., the Lévy function, the Paviani function, $d = 4, 5$) and the motivating pharmaceutical example. That is, based on our investigation, it is prudent to suggest that one could use either of the two correlation functions for such modeling.

Finally, one of the most crucial points in such pharmaceutical applications is that one or more of the factors are often qualitative, whereas $\mathcal{G}$-SELC methodology has been developed for quantitative factors. If there is a quantitative representation of the factors and their levels, our method is useful for qualitative factors as well. Moreover, as argued in Section 2, it is often feasible to rank the monomers (or reagents) based on their intrinsic chemical properties. Although different properties might give rise to different orderings of the monomers, according to our experience, it does not affect the performance of our algorithm significantly. As a guide to practitioners, we recommend using the physicochemical property which is the most important according to the scientists' prior knowledge for assigning the levels of the factors. We have also investigated the impact of relabeling of factor(s) on the proposed approach. The purpose of this investigation is to study the performance of the proposed approach if one (or more) of the factors is qualitative whose levels cannot be ordered. It turns out that the performance of all the three methods (EI, SELC and $\mathcal{G}$-SELC) for the pharmaceutical example discussed in Section 6 is consistent with the results presented in Table 10 (see the Appendix for details). Nonetheless, the proposed methodology can be extended for the qualitative factors. If one or more input factors are qualitative and cannot be treated as quantitative, one can use the correlation function proposed by Qian, Wu and Wu (2008) to model the correlation structure in the Gaussian process model. This is a topic for future research.

## APPENDIX: SIMULATION STUDY FOR THE EFFECT OF RELABELING

Similar to Table 9, Table 11 presents the success rates of the three methodologies for the pharmaceutical example when one of the factors was relabeled. The results are based on 100 simulations. Here, the relabeling "$A \to A + k$" denotes that the new level of $A$ is $\mod(A + k - 1, 5) + 1$. That is, $A \to A + 2$ corresponds to the following relabeling $(1, 2, 3, 4, 5) \longrightarrow (3, 4, 5, 1, 2)$. Similarly, the levels of $B$ and $C$ are relabeled to $\mod(B + k - 1, 34) + 1$ and $\mod(C + k - 1, 241) + 1$, respectively.

**Acknowledgments.** The authors are grateful to the Editor and the Associate Editor for constructive comments on this paper. In addition, we would like to thank Kjell Johnson of Pfizer Global Research and Development, Michigan Laboratories, for helpful comments.

A. Mandal  
Department of Statistics  
University of Georgia  
Athens, Georgia 30602-1952  
USA  
E-mail: amandal@stat.uga.edu

P. Ranjan  
Department of Mathematics and Statistics  
Acadia University  
Wolfville, NS B4P2R6  
Canada  
E-mail: pritam.ranjan@acadiau.ca

C. F. Jeff Wu  
Industrial and Systems Engineering  
Georgia Institute of Technology  
Atlanta, Georgia 30332-0205  
USA  
E-mail: jeffwu@isye.gatech.edu

Table 11
*Simulation results for the pharmaceutical example: success rates of different methods for identifying good compounds, under different relabelings of the factors*

|  | EI | SELC | $\mathcal{G}$-SELC | EI | SELC | $\mathcal{G}$-SELC | EI | SELC | $\mathcal{G}$-SELC | EI | SELC | $\mathcal{G}$-SELC |
|---|---|---|---|---|---|---|---|---|---|---|---|---|
|  | | $A \to A+1$ | | | $A \to A+2$ | | | $A \to A+3$ | | | $A \to A+4$ | |
| Third best | 95 | 100 | 95 | 93 | 100 | 97 | 87 | 100 | 96 | 91 | 100 | 96 |
| Second best | 0 | 0 | 3 | 0 | 0 | 1 | 0 | 0 | 3 | 0 | 0 | 3 |
| Best | 5 | 0 | 2 | 7 | 0 | 2 | 13 | 0 | 1 | 9 | 0 | 1 |
| Total number of good compounds in 100 simulations | 180 | 171 | 300 | 133 | 163 | 270 | 183 | 173 | 281 | 166 | 168 | 295 |
|  | | $B \to B+1$ | | | $B \to B+5$ | | | $B \to B+10$ | | | $B \to B+20$ | |
| Third best | 89 | 100 | 100 | 94 | 99 | 98 | 90 | 100 | 95 | 87 | 100 | 90 |
| Second best | 0 | 0 | 0 | 0 | 1 | 2 | 1 | 0 | 4 | 0 | 0 | 10 |
| Best | 11 | 0 | 0 | 6 | 0 | 0 | 9 | 0 | 1 | 13 | 0 | 0 |
| Total number of good compounds in 100 simulations | 144 | 189 | 264 | 175 | 180 | 260 | 164 | 189 | 275 | 167 | 191 | 305 |
|  | | $C \to C+1$ | | | $C \to C+10$ | | | $C \to C+20$ | | | $C \to C+30$ | |
| Third best | 86 | 100 | 98 | 83 | 100 | 97 | 86 | 99 | 97 | 86 | 100 | 96 |
| Second best | 0 | 0 | 2 | 0 | 0 | 2 | 0 | 1 | 3 | 0 | 0 | 4 |
| Best | 14 | 0 | 0 | 17 | 0 | 1 | 14 | 0 | 0 | 14 | 0 | 0 |
| Total number of good compounds in 100 simulations | 173 | 174 | 277 | 171 | 196 | 288 | 166 | 166 | 314 | 176 | 191 | 296 |